\documentclass[showpacs,showkeys,preprint,prd,nofootinbib,12pt,superscriptaddress]{revtex4-1}

\usepackage{graphicx} 

\usepackage{rotating}
\usepackage{amsmath}
\usepackage{epstopdf}
\usepackage{multirow}
\usepackage{xspace}
\usepackage{rotating}
\usepackage{longtable}
\usepackage{multirow}
\usepackage{placeins}

\usepackage[breaklinks=true]{hyperref}
\hypersetup{
  colorlinks=true,
  linkcolor=blue,
  citecolor=blue,
  urlcolor=blue
}

\makeatletter
\renewcommand*\env@matrix[1][*\c@MaxMatrixCols c]{%
  \hskip -\arraycolsep
  \let\@ifnextchar\new@ifnextchar
  \array{#1}}
\makeatother

\graphicspath{{./fig/}}
\usepackage{array,tabularx,epsfig,mathrsfs,graphicx,rotating}
\usepackage{ifthen}
\usepackage{amsfonts}
\usepackage{subfigure}

\newcommand{\beq}{\begin{equation}}
  \newcommand{\eeq}{\end{equation}}

\chardef\til=126

\def\pt{\ensuremath{p_{\mathrm{T}}}}
\def\ptRes{\ensuremath{(\pt^{\mathrm{reco}}-\pt^{\mathrm{truth}})/\pt^{\mathrm{truth}}}}
\def\genRes{\ensuremath{(\xi^{\mathrm{reco}}-\xi^{\mathrm{truth}})/\xi^{\mathrm{truth}}}}

\begin{document}

\preprint{ANL-HEP-158616}


\date{\today}

\vspace{2.5cm}

\title{
  Automated detector simulation and reconstruction parametrization using machine learning 
}

\author{D. Benjamin}
\affiliation{
  High Energy Physics Division, Argonne National Laboratory,
  9700 S.~Cass Avenue, Argonne, IL 60439, USA 
}
\author{S. Chekanov}
\affiliation{
  High Energy Physics Division, Argonne National Laboratory,
  9700 S.~Cass Avenue, Argonne, IL 60439, USA 
}
\author{W. Hopkins}
\affiliation{
  High Energy Physics Division, Argonne National Laboratory,
  9700 S.~Cass Avenue, Argonne, IL 60439, USA 
}
\author{Y. Li}
\affiliation{
  Computational Science Division, Argonne National Laboratory,
  9700 S.~Cass Avenue, Argonne, IL 60439, USA 
}
\author{J. R. Love}
\affiliation{
  High Energy Physics Division, Argonne National Laboratory,
  9700 S.~Cass Avenue, Argonne, IL 60439, USA 
}

\begin{abstract}
Rapidly applying the effects of detector response to physics objects (e.g. electrons, muons, showers of particles) is essential in high energy physics. Currently available tools for the transformation from truth-level physics objects to reconstructed detector-level physics objects involve manually defining resolution functions. These resolution functions are typically derived in bins of variables that are correlated with the resolution (e.g. pseudorapidity and transverse momentum). This process is time consuming, requires manual updates when detector conditions change, and can miss important correlations. Machine learning offers a way to automate the process of building these truth-to-reconstructed object transformations and can capture complex correlation for any given set of input variables. Such machine learning algorithms, with sufficient optimization, could have a wide range of applications: improving phenomenological studies by using a better detector representation, allowing for more efficient production of Geant4 simulation by only simulating events within an interesting part of phase space, and studies on future experimental sensitivity to new physics.
\end{abstract}

\maketitle

\section{Introduction}

A cornerstone of particle collision experiments is the Monte Carlo (MC) simulation of physics processes resulting from collisions of high-energy particles, followed by the simulation of detector responses and object reconstruction. The MC simulation produces objects (jets, electrons, muons, etc) with properties (four momenta, particle types) which entirely depend on the physics processes occurring. These objects are commonly referred to as ``truth'' objects. These objects are altered by interactions with the detector and are reconstructed with experimental algorithms. Such objects, that have undergone a transformation due to detector interactions and reconstruction will be referred to as ``reco'' objects in this paper.

With the increased complexity of high-energy collider experiments, such as those at the Large Hadron Collider (LHC), the detector simulations become increasing complex and time consuming. Parameterized detector simulations, such as Delphes~\cite{deFavereau:2013fsa}, have been proven to be a vital tool for physics performance and phenomological studies (i.e. to estimate the sensitivity of an experiment to a new physics model). An approximation of the detector responses and experimental object reconstruction can, however, also be performed by a neural network (NN) trained using the Geant4-based simulations that have gone through an experiment's reconstruction algorithm. Such an NN could then computationally rapidly transform truth MC objects (jets and other identified particles) to objects modified by a detector and experimental reconstruction algorithms.  

The main advantage of a detector parametrization based on machine learning (ML), as compared to a manually-constructed analytic parametrization such as Delphes, is that a neural network can automatically learn the features introduced by detailed full simulations avoiding the need to handcraft parameters to represent resolutions and inefficiencies. An NN trained using realistic detector simulation could memorize the transformation from the truth to the reco quantities without dedicated studies of resolution functions. Another advantage is that the NN approach can introduce a complex interdependence of variables which is currently difficult to implement in parameterized simulations. Finally, since the underlying libraries used for ML (e.g. Keras~\cite{chollet2015keras}, pyTorch~\cite{NEURIPS2019_9015}, etc) are optimized for a wide range of hardware, an NN-based truth-to-reco transformation would be able to run efficiently on heterogeneous hardware resources (resources that use a varied set of processors such as GPUs and CPUs).

As a first step towards parameterized detector simulations with ML, it is instructive to investigate how a transformation from the truth to reco objects can be performed, leaving aside the question of introducing objects that are created by misreconstructions or objects that are lost due to inefficiencies.

\section{Traditional parameterized fast simulations}

In abstract terms, a typical variable $\xi_i^{\mathrm{reco}}$ that characterizes a reconstructed particle/jet, such as transverse momentum (\pt$^{\mathrm{reco}}$) or pseudorapidity ($\eta^{\mathrm{reco}}$), can be viewed as the result of a multivariate transform, $F$, of the original variable $\xi_1^{\mathrm{truth}}$ at truth level:

$$
\xi_1^{\mathrm{reco}} = F (\xi_1^{\mathrm{truth}}, \xi_2^{\mathrm{truth}}, \xi_3^{\mathrm{truth}}, ...\xi_{\mathrm{N}}^{\mathrm{truth}}).
$$
Generally, such a transform  depends on several other variables $\xi_2^{\mathrm{truth}}$ ..  $\xi_{\mathrm{N}}^{\mathrm{truth}}$ characterizing this (or other) objects at truth level. For example, the extent at which jet transverse momentum, \pt\ is modified by a detector depends on the original truth-level transverse momentum ($\xi_1^{\mathrm{truth}}=p_T^{\mathrm{truth}}$), pseudorapidity ($\xi_2^{\mathrm{truth}}=\eta^{\mathrm{truth}}$), and other effects that can be inferred from truth quantities. Similarly, if particular detector regions in the azimuthal angle ($\phi$) have low efficiency, this would introduce an additional dependence of this transform on $\phi$.

Typical parameterized simulations ignore the full range of correlations between the truth-level variables. In most cases, the above transform is reduced to a single variable, or two (as in the case of Delphes simulations where the energy resolution of clusters depends on the original energies of particles and their positions in $\eta$). In order to take into account correlations between multiple parameters characterizing transformations to reconstruction objects, a grid in the hypercube with the dimension $N_{\mathrm{b}}^{\mathrm{N}}$, where $N_{\mathrm{b}}$ is the number of histogram bins for the distributions \genRes, representing the ``resolution'', must be created. This methodology results in a large number of histograms when there are many correlated variables that affect the resolution.

It should be pointed out that the calculation speed for parameterized simulations of one variable that depends on $N$ other variables at the truth level is proportional to $N_{\mathrm{b}}^{\mathrm{N}}$ since each object at the truth level should be placed inside the grid defined by $N_{\mathrm{b}}$ bins. Therefore, complex parameterisations of resolutions and efficiencies for $N>2$ becomes computationally intensive. 

\section{Jet truth-to-reco transformation with ML}

\begin{figure}[h]
  \includegraphics[width=0.8\textwidth]{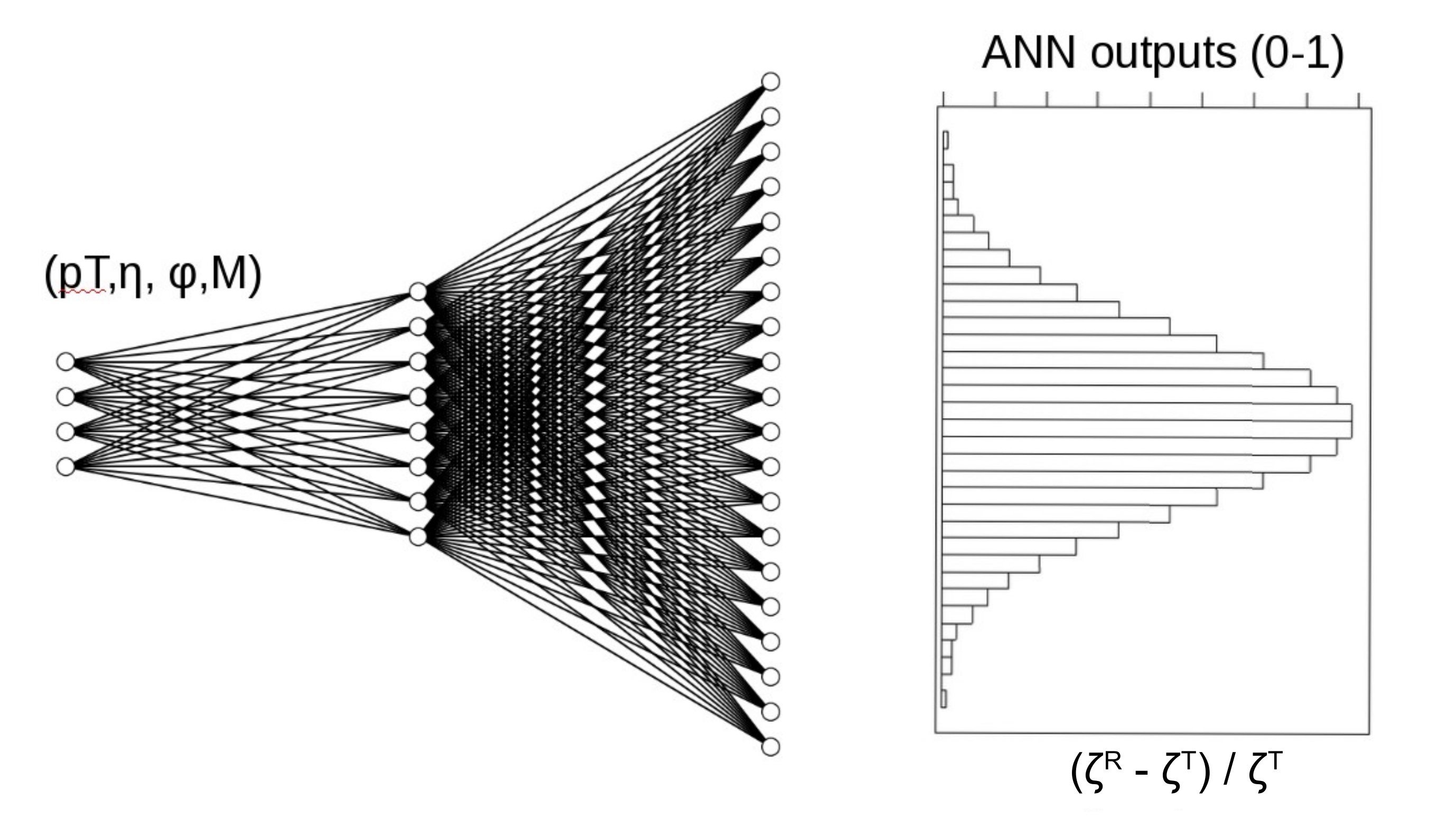}
  \caption{A schematic representation of the NN architecture for modelling the detector response and affect of reconstruction algorithms on truth-level input variables. The output nodes of this NN represent a binned PDF for the resolution of single variable, e.g. \ptRes, while other variables such $\eta$, $\phi$, mass ($m$) are auxiliary variables that affect \ptRes.}
  \label{ann_example}
\end{figure}

To test the viability of using ML to transform truth objects to reco objects, we studied the truth-to-reco transformation for jets. Jet truth-level quantities, such as jet $\pt$, $\eta$, $\phi$ and jet mass ($m$) are used as training inputs to an NN while the output
is an array of nodes that represent the binned probability density function (PDF) of the resolution for a single variable (such as jet \pt). Additional input can consist of any variable that can influence the resolution of a jet, such as jet flavor at the truth level, jet radius, etc. Figure~\ref{ann_example} shows a schematic representation of the NN architecture for modelling the detector response for a single output variable. The aim is to have the NN learn the shape of the resolution PDF, for example for the $\pt$, depending on other input variables such as the $\eta$ of the object. A binned output (multi-categorization) was used so that the precision of the resolution PDF modelling can be chosen.

\section{Monte Carlo simulated event samples}
\begin{figure}[h]
  \includegraphics[width=0.48\textwidth]{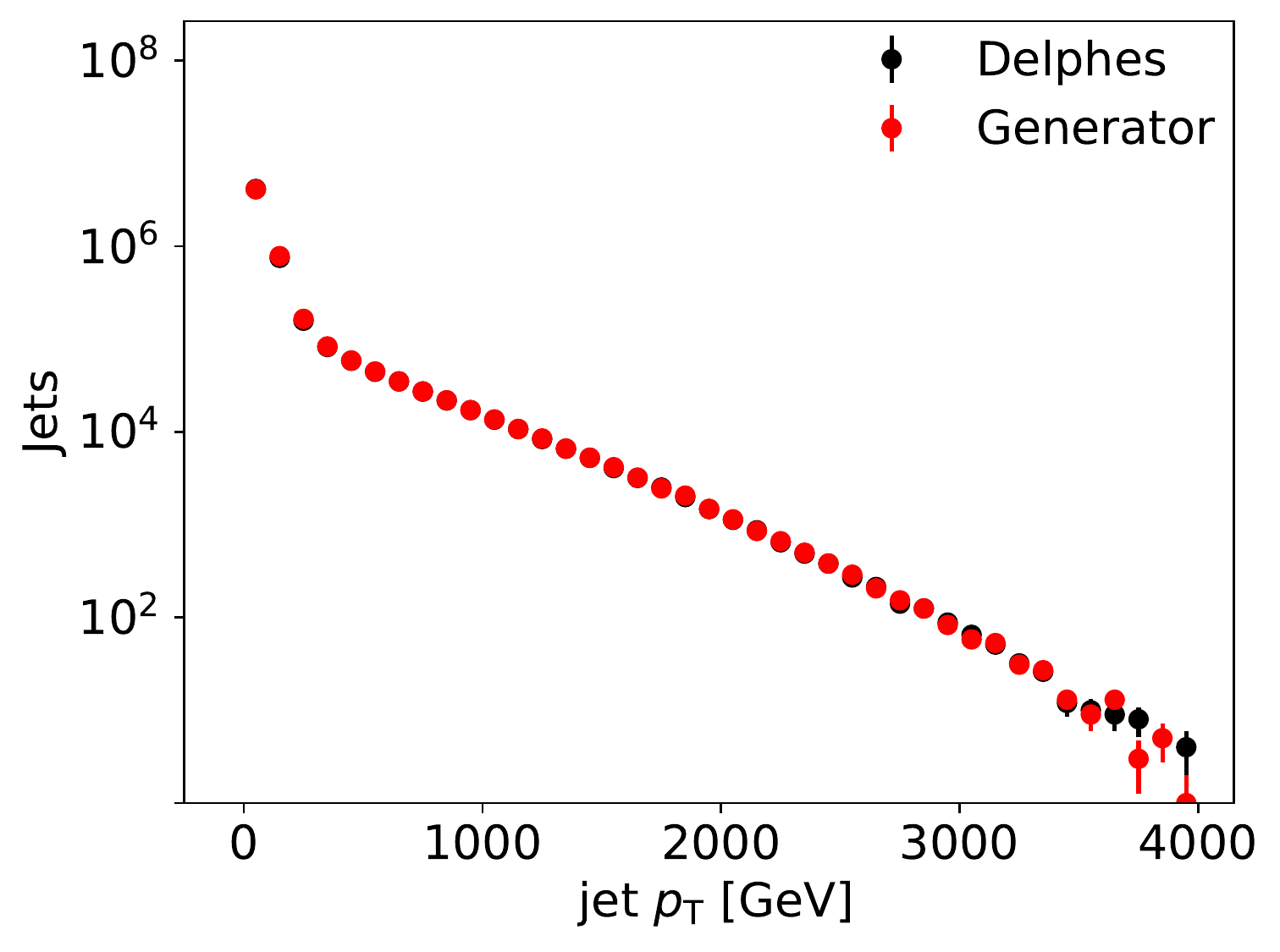}
  \includegraphics[width=0.48\textwidth]{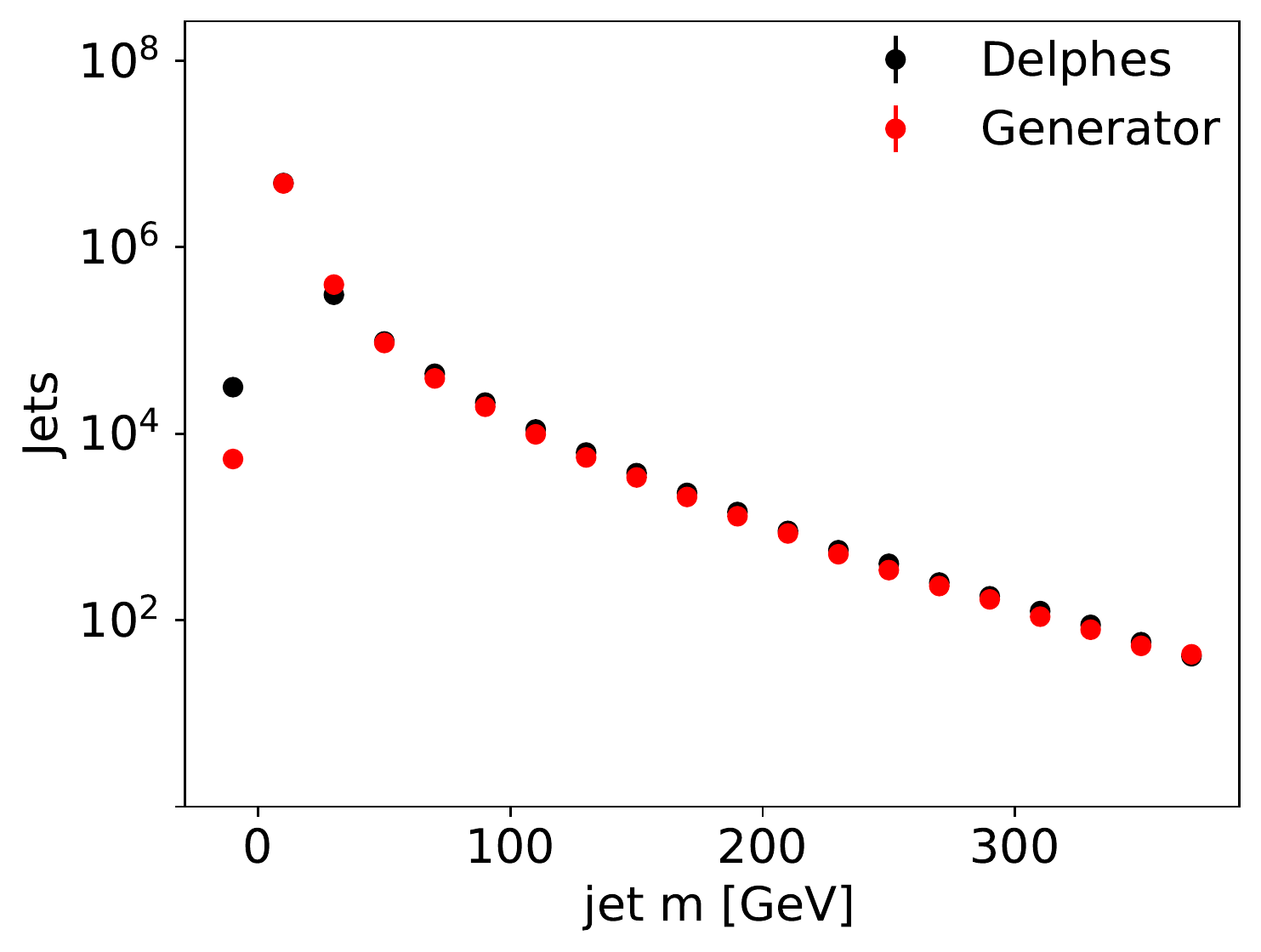}\\
  \includegraphics[width=0.48\textwidth]{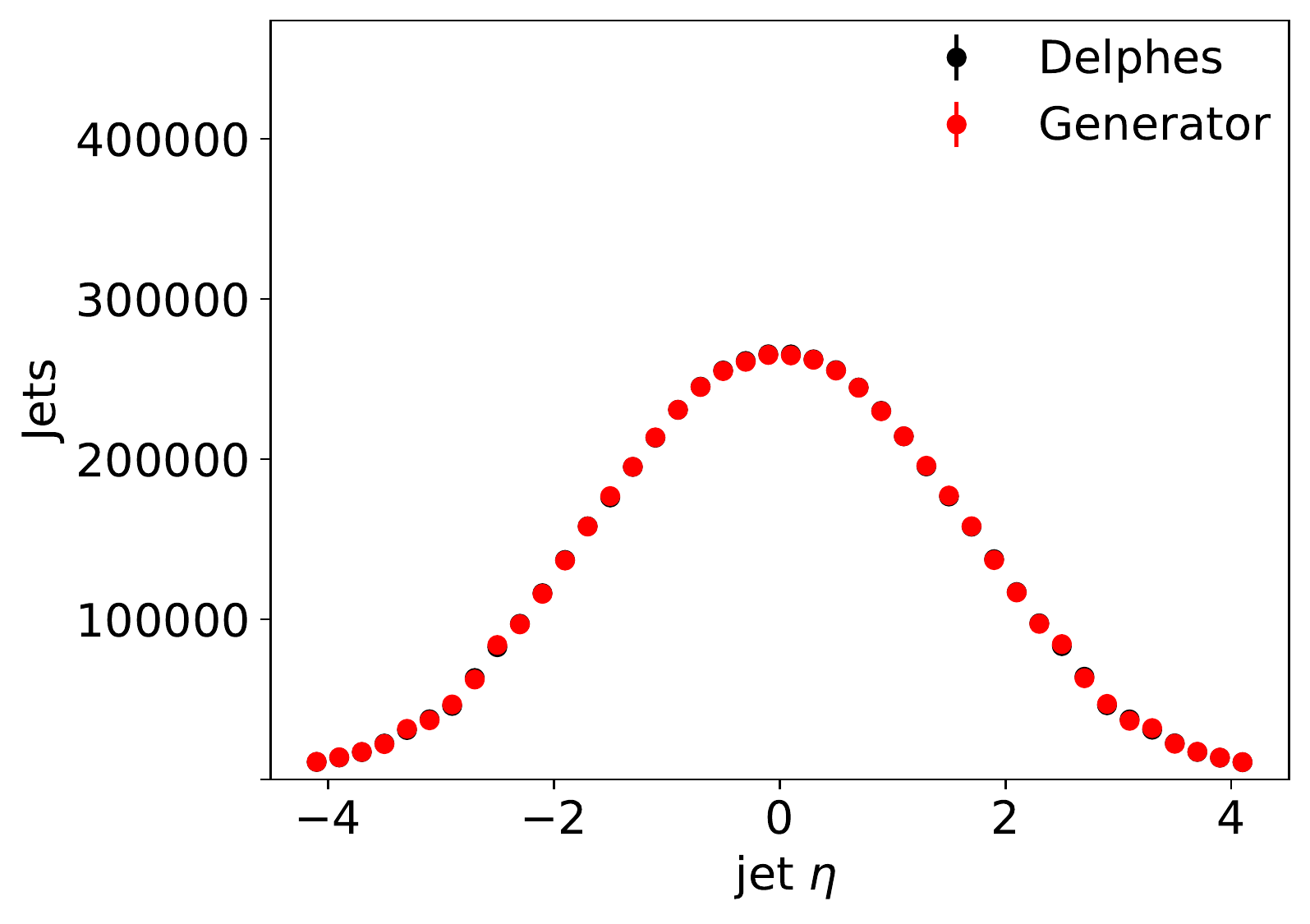}
  \includegraphics[width=0.48\textwidth]{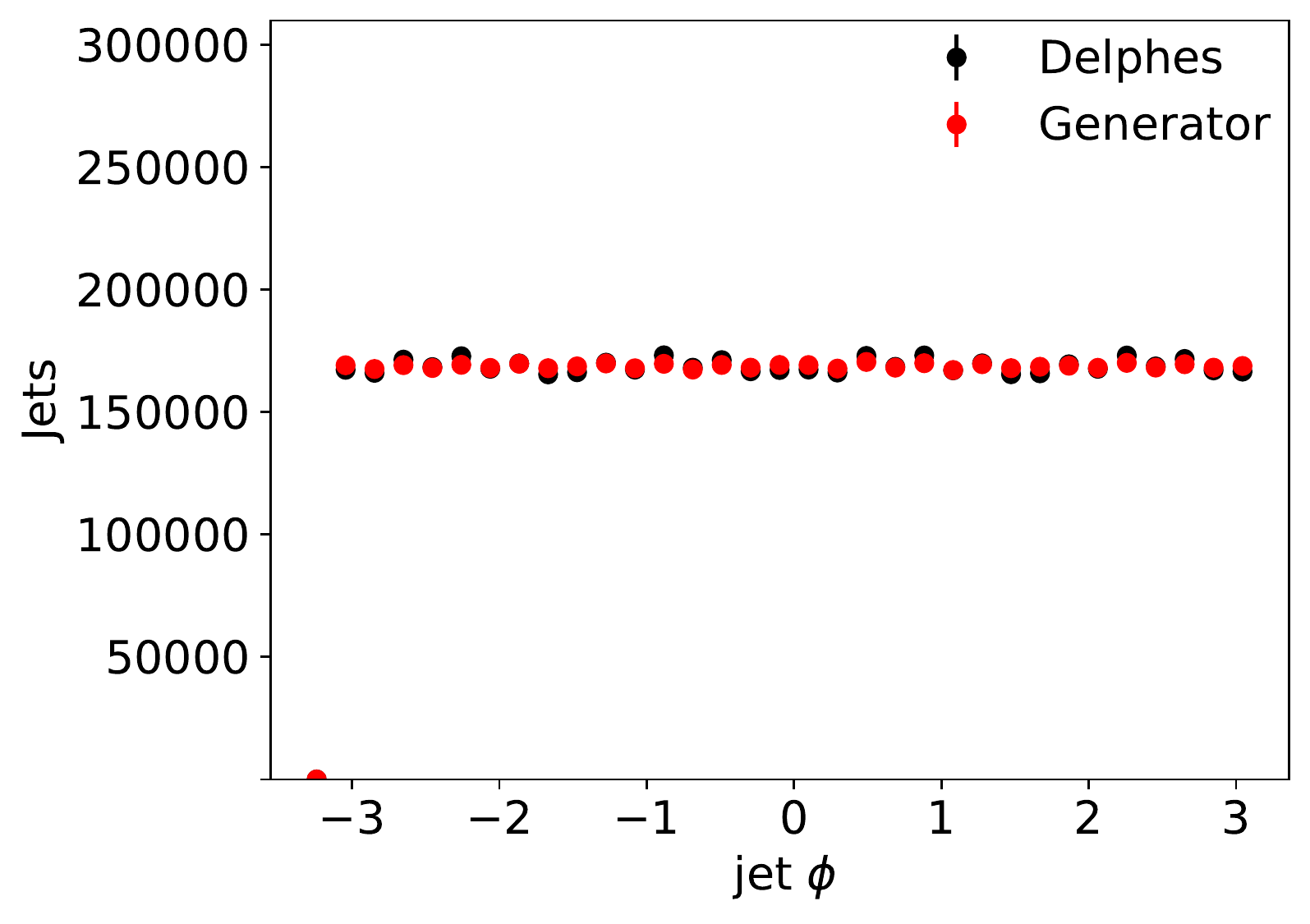}
  \caption{Distributions for input variables for truth (red) and reco quantities (black).}
  \label{fig:nnInputsPrescaling}
\end{figure}

\FloatBarrier
Monte Carlo events used for this analysis were produced using the Madgraph generator~\cite{Alwall:2014hca}. The simulated processes are a combination of equal event samples with  top pair production ($t\bar{t}$) and photons produced in association with jets ($\gamma$+jets), which give a high rate of jets in different environments. 
Hadronic jets were reconstructed with the {\sc FastJet} package~\cite{Fastjet} using the anti-$k_t$ algorithm \cite{Cacciari:2008gp} with a distance parameter of 0.4. The detector simulation was performed with the Delphes package with a detector geometry which is similar to the ATLAS geometry. 
The event samples used for the following study are available from the HepSim database~\cite{Chekanov:2014fga}. In this paper, only the transformation of \pt\ from truth jets (which have truth particle constituents) to reconstructed jets (which have calorimeter cell constituent) was performed. However, the methodology should be object and parameter agnostic. Only truth jets which are matched to a reconstructed Delphes jet are considered in this study. For the matching criteria the reconstructed jet that has the smallest $\Delta R=\sqrt{\Delta\phi^2+\Delta\eta^2}$, where $\Delta\phi=\phi^{\text{truth}}-\phi^{\text{reco}}$ and $\Delta\eta=\eta^{\text{truth}}-\eta^{\text{reco}}$, with respect to the truth jet is chosen. If this minimum $\Delta R$ is greater than 0.2, the truth jet is discarded. No other requirements are made on truth and reconstructed Delphes jets other than the $\pt>15$ GeV requirement in Delphes. Only matched jets are used since the aim of the study is to test whether an NN can learn changes in detector resolution as a function of kinematic properties of the jet (e.g. $\pt$, $\eta$, $\phi$, $m$). 
The final number of training jets used is two million while 500,000 jets were used as an independent test sample. The distributions of quantities used as the input for the NN, \pt, $\eta$\, $\phi$, $m$, are shown in Figure~\ref{fig:nnInputsPrescaling}.

To facilitate gradient descent in all direction of the input variable space, the input variables are scaled to be in the range [0,1]. This avoids the \pt\ and the mass from having a disproportional affect on the training of the NN. The output variable, \ptRes, is also scaled to have values between 0 and 1. Only jets that 
are within the 1$^{\mathrm{st}}$ and 99$^{\mathrm{th}}$ percentile of the \ptRes\ distribution are considered.

\begin{figure}[h]
  \includegraphics[width=0.48\textwidth]{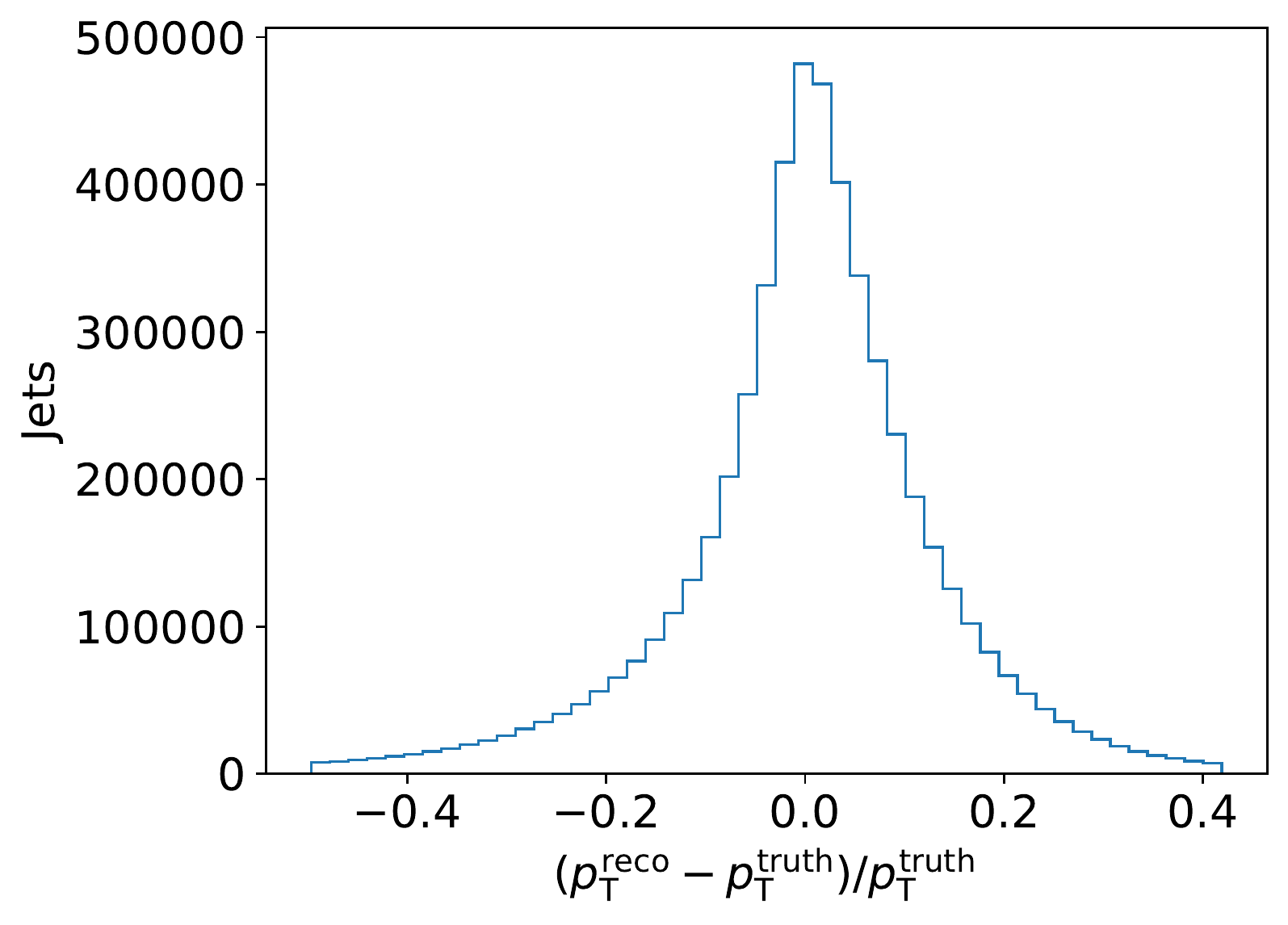}
  \caption{Resolution of the \pt\ (relative differences between truth and reco \pt).}
  \label{fig:deltaTarget}
\end{figure}

\section{Neural network structures}

An NN is trained with four input parameters, the scaled \pt, $\eta$, $\phi$, and $m$, and consist of five layers with 100 nodes each and with each node having a rectifier linear unit (ReLu) activation function. Several output layer configurations with 100, 200, 300, 400, and 500 output nodes were tested, all with a softmax activation function. The configuration with 400 output nodes resulted in the best performance, measured by how well the NN could mimic the Delphes \pt\ spectrum and resolution (see below for details), with the least number of total NN parameters. 

In an attempt to optimize the NN training, several batch-size and number-of-epoch combinations were used in an attempt to improve the sensitivity to a small subsample (the forward jets) of the training sample. The number of backpropagations ($N_{bp}$) were held constant by keeping the ratio of the number of epochs ($N_e$) and batch size ($N_b$) constant since $N_{bp}=\frac{N_t}{N_b}N_e$ where $N_t$ is the number of training jets. Batch size and number of epochs of 5, 10, 20, 100, 200, 1000 were tested resulting in similar performance of the NN.

Finally, the NN is trained using the Adam~\cite{adam} optimizer with a learning rate of $10^{-4}$ and is implemented using Keras with a TensorFlow~\cite{tensorflow2015-whitepaper} backend.

\section{Results}
\FloatBarrier

\begin{figure}[htb]
  \subfigure[]{\includegraphics[width=0.48\textwidth]{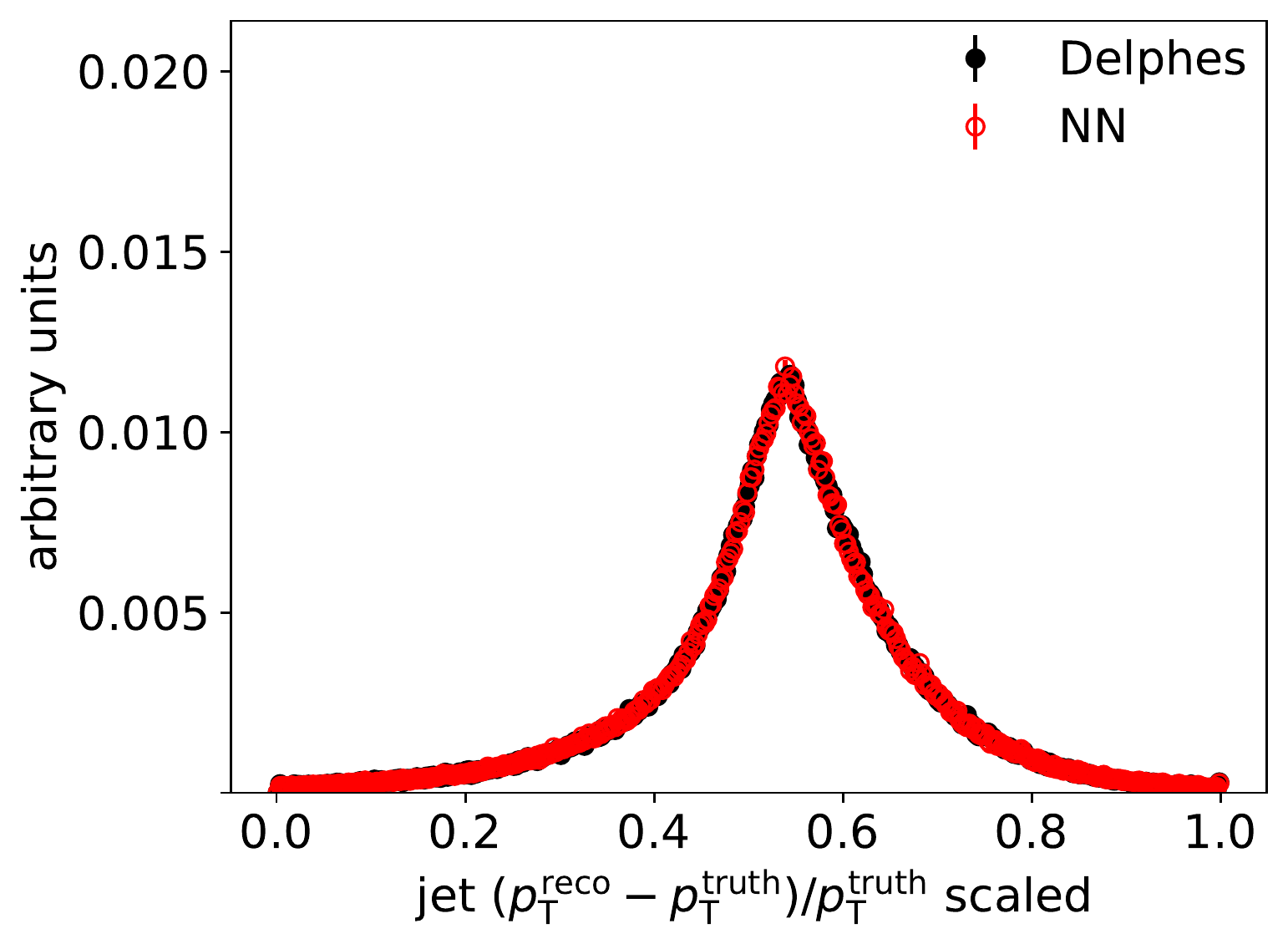}}
  \subfigure[]{\includegraphics[width=0.48\textwidth]{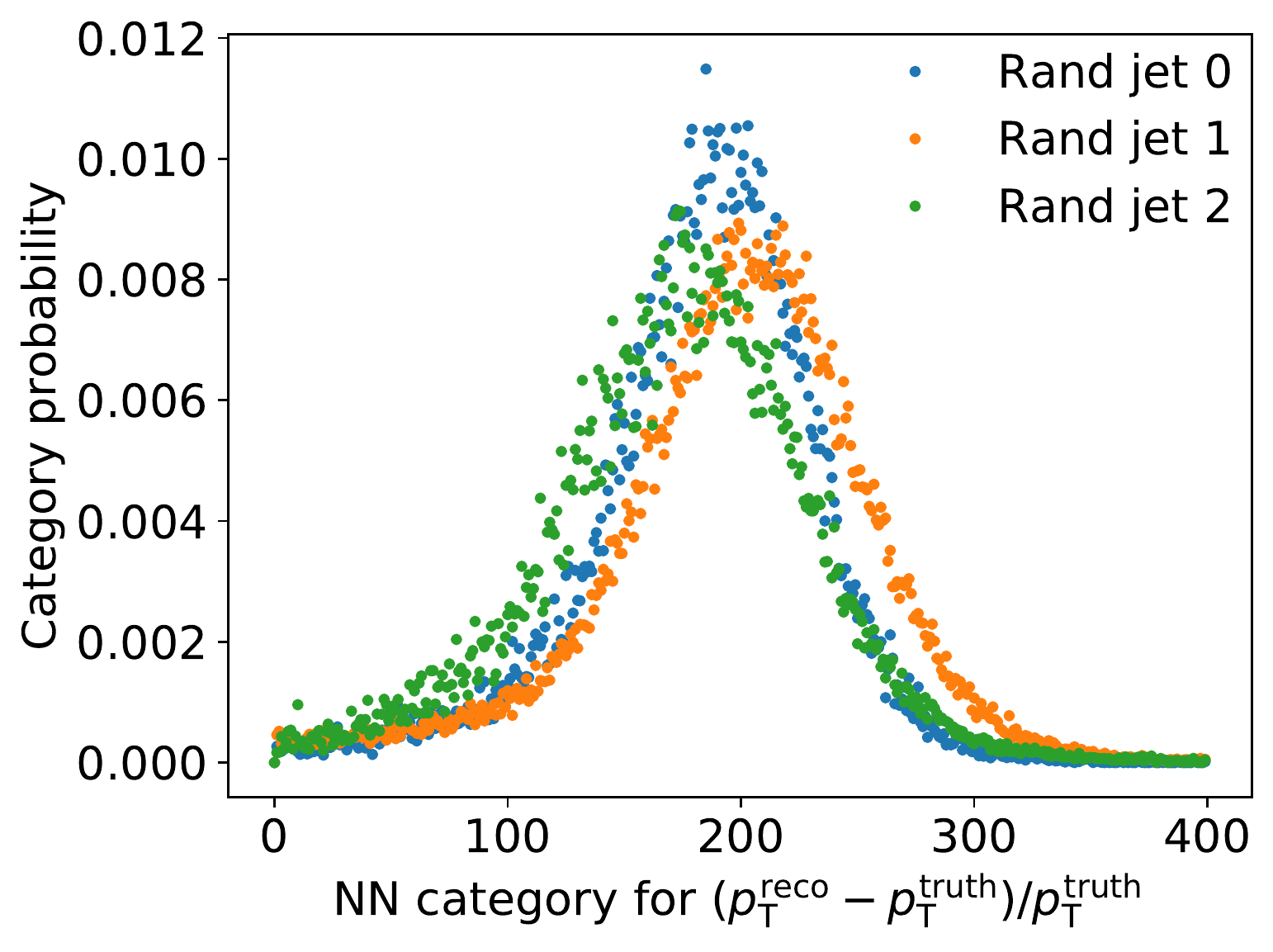}}
  \caption{NN-generated jet \ptRes\ compared to Delphes reco jet \ptRes\ (a). Representative values of the NN output after training for three randomly selected truth jets which have different input values (b).}
  \label{fig:PDFComp}
\end{figure}

After the NN has been trained to learn the PDF of \ptRes, the resulting learned PDF is compared to the Delphes PDF using the test sample in Figure~\ref{fig:PDFComp}a. Good agreement is observed between the Delphes and NN PDFs, showing that the NN has learned the bulk distribution.

\begin{figure}[htb]
  \subfigure[]{\includegraphics[width=0.42\textwidth]{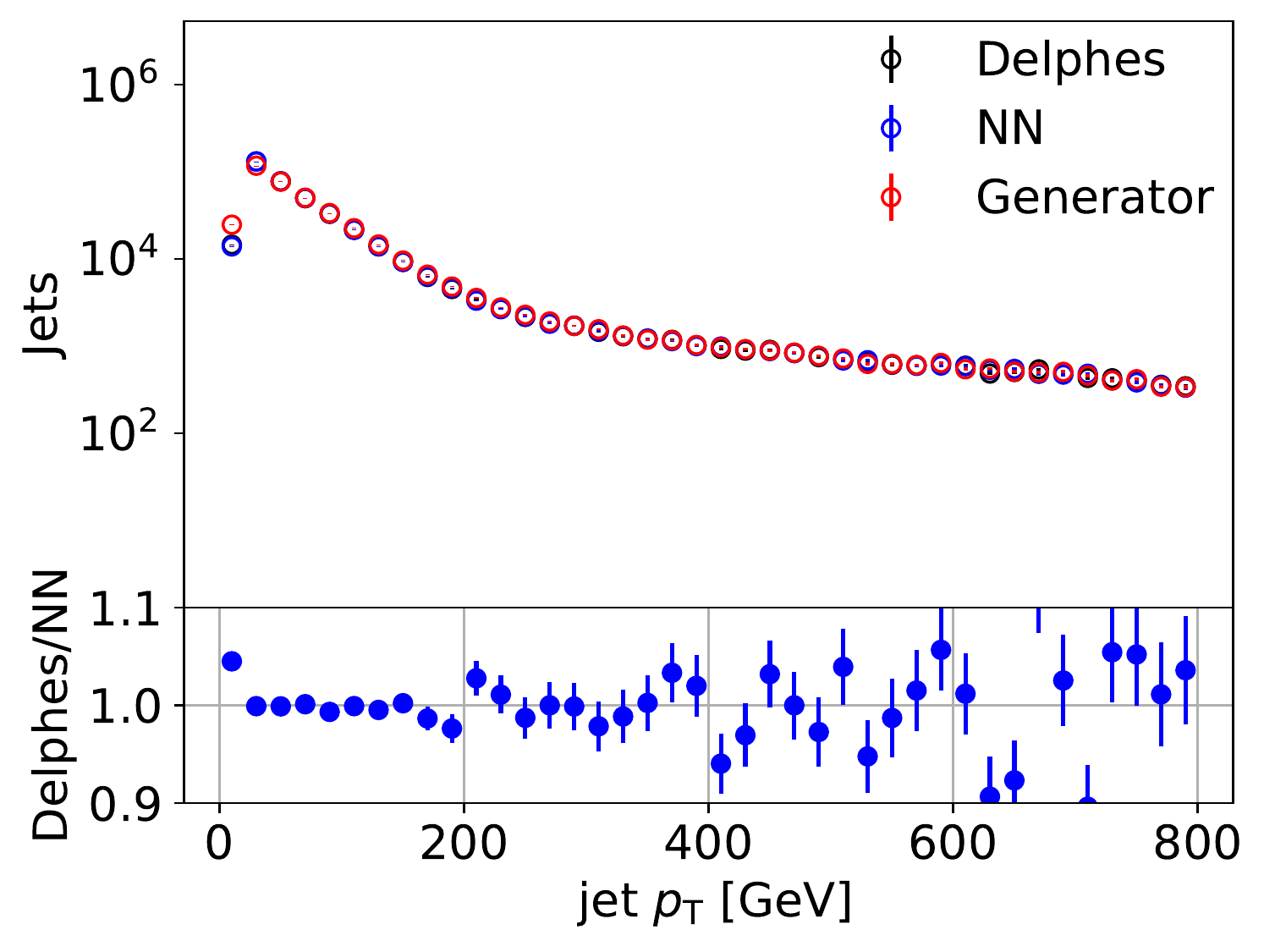}}
  \subfigure[]{\includegraphics[width=0.48\textwidth]{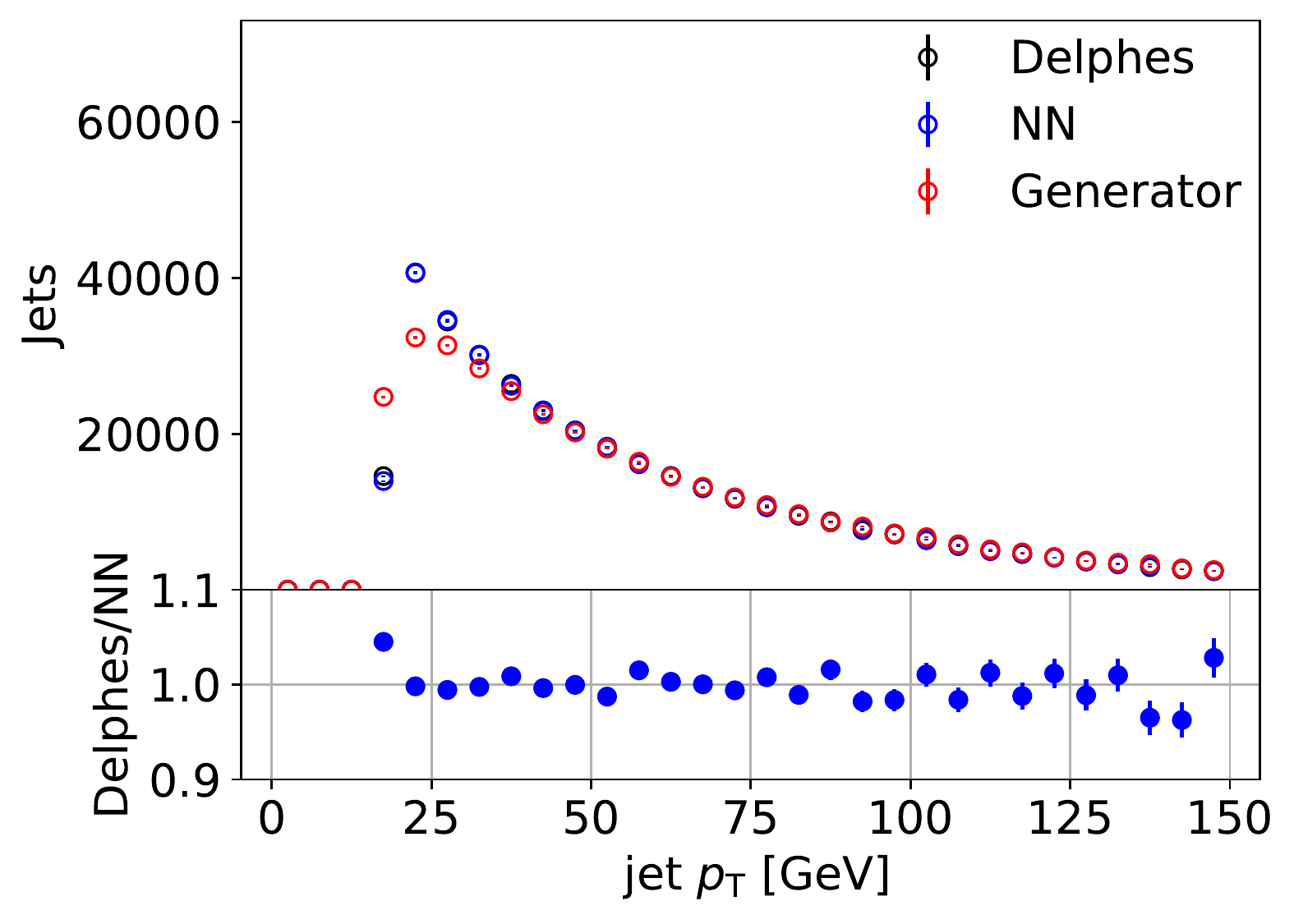}}
  \caption{Delphes and NN-generated jet \pt\ disitributions for a wide (a) and narrow (b) \pt\ range.}
  \label{fig:pTNNVsDelphes}
\end{figure}

The NN output represents a binned PDF for each jet based on its input parameters (i.e. \pt, $\phi$, $\eta$, and $m$). The PDFs for a set of three randomly selected jets are shown in Figure~\ref{fig:PDFComp}b which features shapes expected for typical resolution function with variations due to changes in jet input parameters. These PDFs are then randomly sampled to produce an NN jet that mimics the reco jet. A comparison of the NN-generated and Delphes jet \pt\ distribution for the test sample is shown in Figure~\ref{fig:pTNNVsDelphes}. The NN reproduces the jet \pt\ distribution of Delphes within 5\% for reconstructed jets with $\pt>20$ GeV.

\begin{figure}[htb]
  \includegraphics[width=0.48\textwidth]{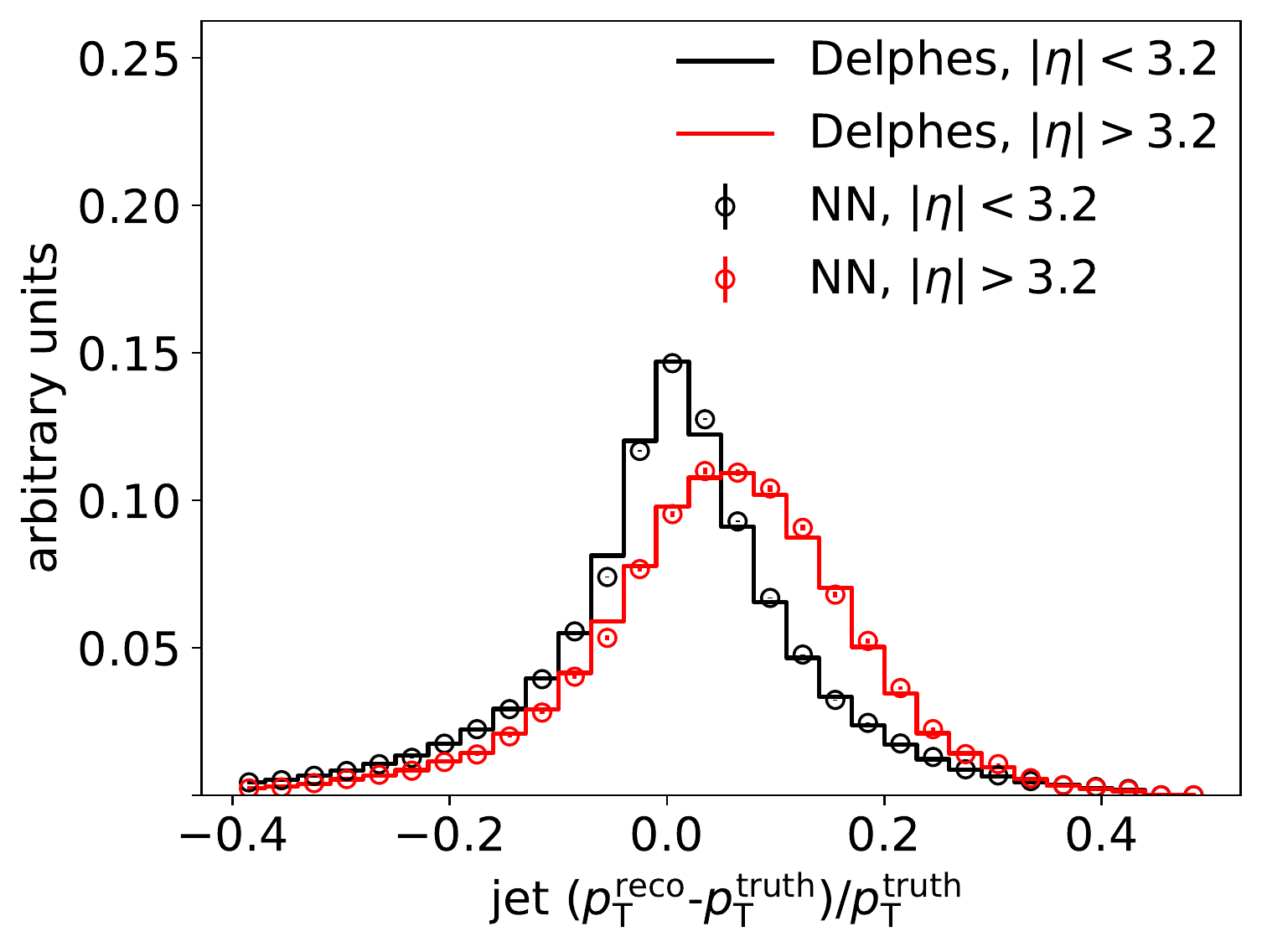}
  \caption{Jet \pt\ resolution for the training sample for both the central and forward region. }
  \label{fig:nnRes}
\end{figure}

To test whether the NN learned correlations between input parameters and the \pt\ resolution, the jets were divided into central ($|\eta|<3.2$) and forward ($|\eta|>3.2$) jets. The \pt\ resolution is then compared between the two regions for both the Delphes jets as well as the NN-generated jets. These two regions in the detector simulation have different calorimeter responses which results in different jet \pt\ resolutions in these two $|\eta|$ regions. The resulting resolutions for both regions are shown in Figure~\ref{fig:nnRes} using the training sample. The training sample was chosen for this comparison because forward jets make up a small subsample of all jets, as can be seen in Figure~\ref{fig:nnInputsPrescaling}.

\begin{figure}[htb]
  \subfigure[]{\includegraphics[width=0.49\textwidth]{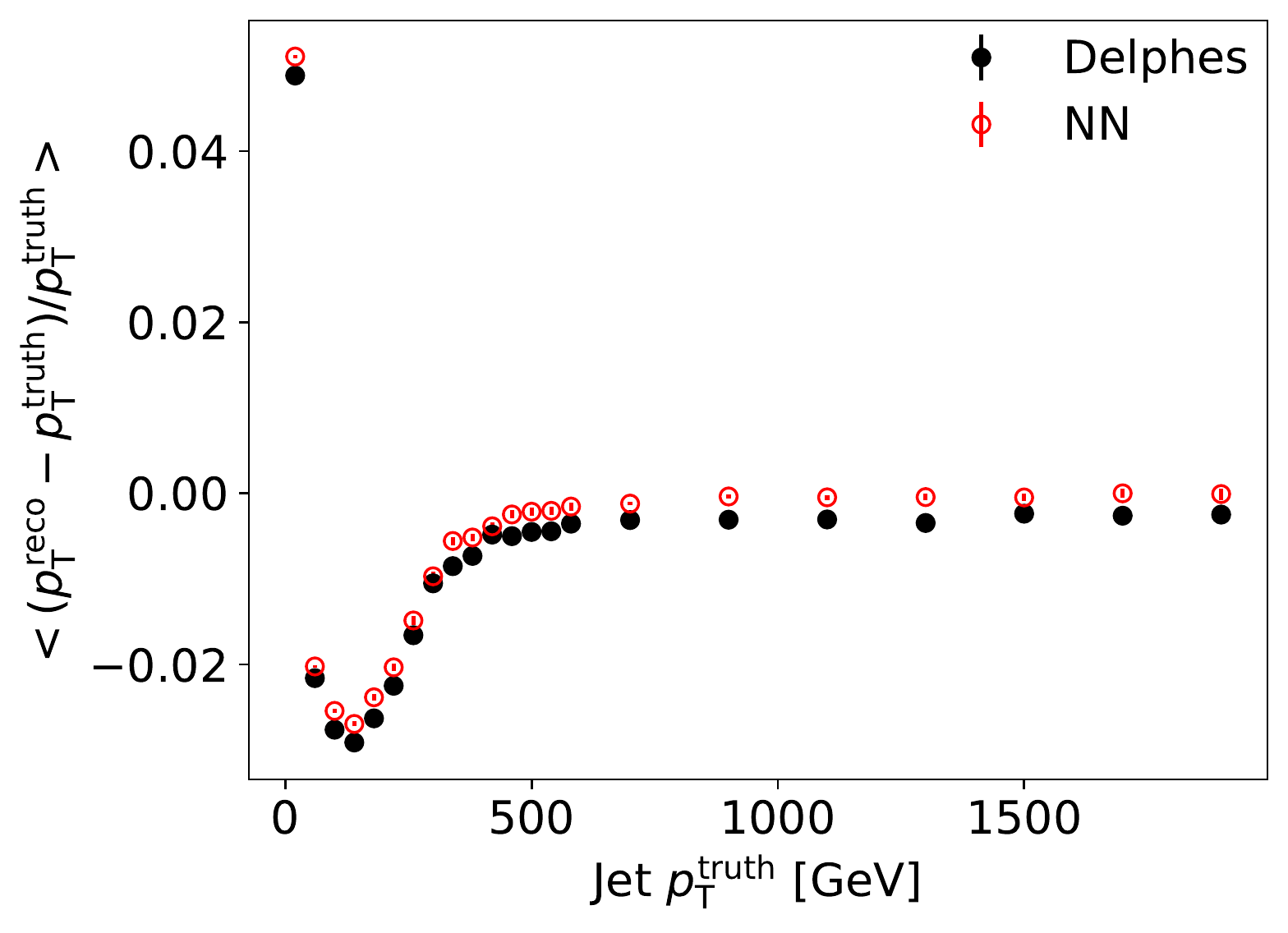}}
  \subfigure[]{\includegraphics[width=0.48\textwidth]{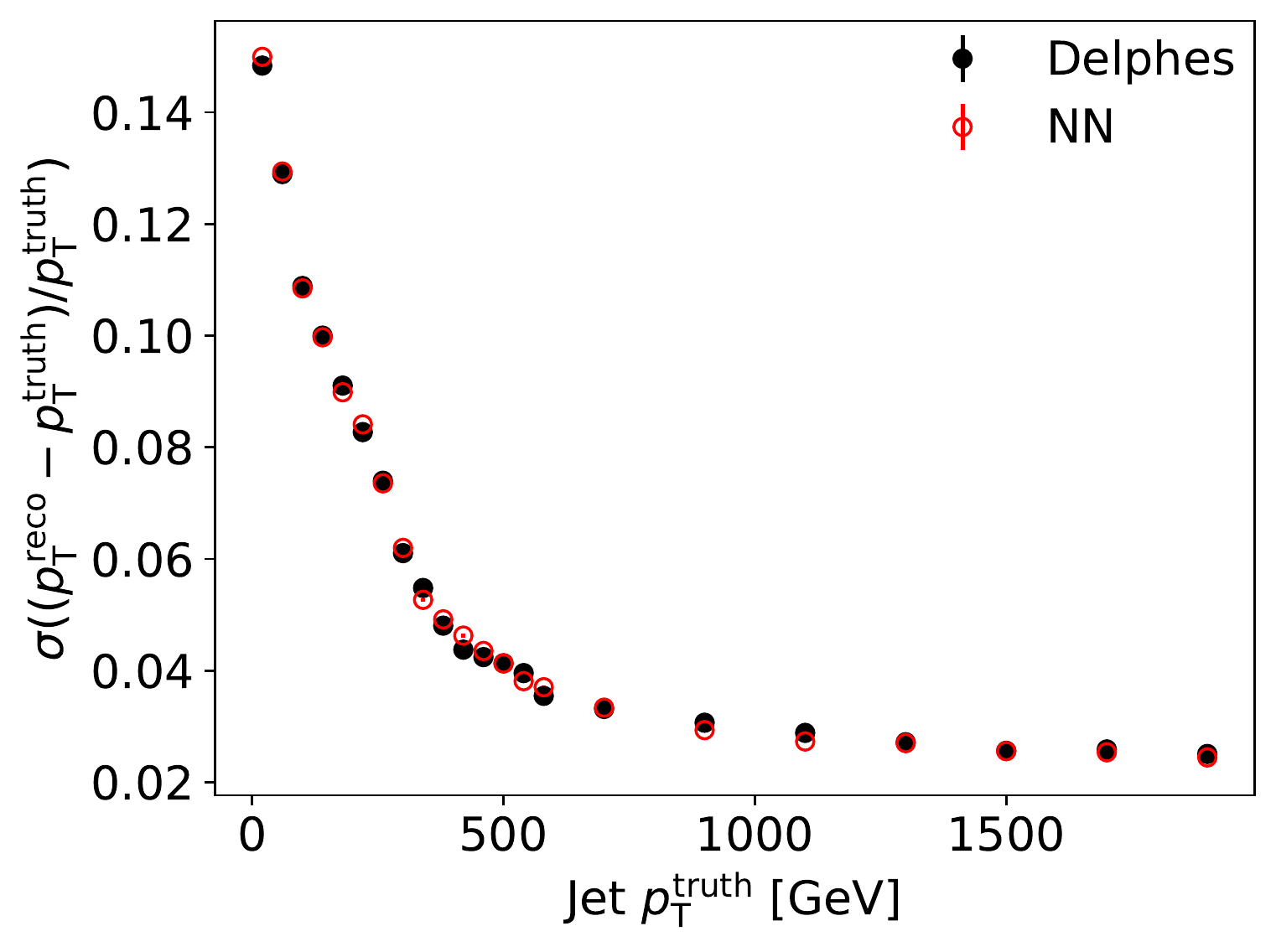}}
  \caption{The mean (a) and standard deviation of the jet \pt\ resolution for Delphes and NN-generated as a function truth jet \pt.}
  \label{fig:resVsPt}
\end{figure}

The mean and standard deviation of the resolution (demonstrated in Figure~\ref{fig:PDFComp}) as a function of \pt\ is shown in Figure~\ref{fig:resVsPt}. The mean of the resolution for the NN is systematically higher than the resolution for Delphes but this effect is small when considering the width of the resolution. The standard deviation of the resolutions, however, are the same for the NN and Delphes across the \pt\ range showing that the NN accurately predicts the resolutions for a large range in \pt. 

In order to produce an NN with optimal performance and a minimal amount of parameters, a genetic algorithm (GA) is used to optimize the NN hyperparameters. The number of layers, number of neurons in each layer, and choice of the learning rate are scanned to find the optimal configuration. The GA we utilize is an evolutionary algorithm that mimics the process of natural selection and which was previously used in the determination of the parameters of a complex force field~\cite{doi:10.1021/acs.jctc.7b00521,doi:10.1021/acs.jctc.6b00432}. Due to the small scale of our problem, the initial configuration was found to be equivalent to the optimal configuration found by the GA.

\section{Conclusion}
A truth-level to reconstruction-level transformation using a multi-categorizing NN is presented. This approach does not require the determination of analytic resolution functions since an NN can automatically learn the resolutions during the training procedure. The NN implementation presented effectively learned the truth-to-reconstruction transformation without requiring manual binning to capture the differences in resolutions of particular subsamples (i.e. central and forward jets). The automatic learning of correlations between the input variables and the resolution is one of the attractive features of using an ML-based transformation, allowing for rapid deployment of detector parametrizations.

Additional improvements could probably be made by including more information about the objects (e.g. whether a $b$-quark is present in a jet, kinematic information from other objects in the event) making this method more robust. This method should be easily extendable to additional reconstructed quantities and could be used to model the ATLAS and CMS detector. The method described in this paper allows for automated detector parametrization which can facilitate phenomological studies, efficient truth event selection, and upgrade studies.

\section*{Acknowledgments}
We gratefully acknowledge the computing resources provided on a
high-performance computing cluster operated by the
Laboratory Computing Resource Center at Argonne National Laboratory.
The submitted manuscript has been created by UChicago Argonne, LLC,
Operator of Argonne National Laboratory (“Argonne”). Argonne, a U.S.
Department of Energy Office of Science laboratory, is operated under
Contract No. DE-AC02-06CH11357. The U.S. Government retains for itself,
and others acting on its behalf, a paid-up nonexclusive, irrevocable
worldwide license in said article to reproduce, prepare derivative works,
distribute copies to the public, and perform publicly and display
publicly, by or on behalf of the Government.
The Department of Energy will provide public access to these results of
federally sponsored research in accordance with the
DOE Public Access Plan.
\url{http://energy.gov/downloads/doe-public-access-plan}. Argonne
National Laboratory’s work was
funded by the U.S. Department of Energy, Office of High Energy Physics
under contract DE-AC02-06CH11357.

\bibliographystyle{unsrt}   
\bibliography{main}  

\end{document}